\documentclass{article}

\usepackage{amsmath}
\usepackage{amsfonts}
\usepackage{graphicx}
\usepackage{caption}
\usepackage{subcaption}

\usepackage{authblk}
\usepackage{amssymb}

\usepackage{epstopdf}
\usepackage{epsfig}
\usepackage{hyperref}

\usepackage{color}

\date{\today}

\begin{document}

\title{Scalarization of neutron stars with realistic equations of state}

\author{Zahra Altaha Motahar}
\author{Jose Luis Bl\'azquez-Salcedo}
\author{Burkhard Kleihaus}
\author{Jutta Kunz}

\vspace{0.5truecm}
\affil{Institut f\"ur  Physik, Universit\"at Oldenburg, Postfach 2503,
D-26111 Oldenburg, Germany}

\vspace{0.5truecm}

\vspace{0.5truecm}

\vspace{0.5truecm}
\date{
\today}

\maketitle

\begin{abstract}
We consider the effect of scalarization
on static and slowly rotating neutron stars for a wide variety
of realistic equations of state, including pure nuclear matter,
nuclear matter with hyperons, hybrid nuclear and quark matter,
and pure quark matter.
We analyze the onset of scalarization, presenting a universal
relation for the critical coupling parameter versus compactness.
We find that the onset and the magnitude of the scalarization are 
strongly correlated with the value of the gravitational potential
(the metric component $g_{tt}$) at the center of the star.
We also consider the moment-of-inertia--compactness relations
and confirm universality for the nuclear matter, hyperon and hybrid
equations of state.
\end{abstract}

\section{Introduction}

Due to their compactness and high density, neutron stars
represent ideal laboratories to test alternative theories of gravity
\cite{Will:2005va,Capozziello:2010zz,Berti:2015itd}. At the same time,
neutron stars are important probes to better understand the properties
of matter under extreme conditions. 
Currently, a large number of equations of state describing
high density matter still seem to be observationally viable
(see e.g.~\cite{Ozel:2016oaf,Lattimer:2012nd}).

Much recent progress in determining the properties of neutron stars,
such as their masses and radii, has been achieved
by exploiting a variety of observational techniques,
including, in particular radio observations of pulsars and 
X-ray observations of neutron stars in low-mass X-ray binaries.
Neutron stars represent also a major focus of the gravitational 
wave detector Advanced LIGO, where the detection of 
neutron star--neutron star and neutron star--black hole mergers 
is expected (see, e.g.,~\cite{Faber:2012rw}).

Certainly, most studies of neutron stars have been based 
on general relativity (GR). However, it is essential to
study the properties of neutron stars also in currently
viable alternative theories of gravity \cite{Berti:2015itd},
where scalar-tensor theory (STT) represents a most prominent
example \cite{Brans:1961sx,Damour:1992we,Fujii:2003}. 
In particular, STT represents a natural generalization of GR,
where one or more scalar fields are included as additional mediators
of the gravitational force.

In the context of neutron stars in STT, an interesting phenomenon 
called spontaneous scalarization (in analogy to spontaneous magnetization)
has been found by Damour and Esposito-Far\`ese \cite{Damour:1993hw,Damour:1996ke}.
Here besides the GR solutions with a vanishing scalar field,
new configurations with a nontrivial scalar field can arise,
because the scalar field nonlinearities can intensify the
attractive nature of the scalar field interactions,
when there are suitable conditions within the star.

The phenomenon of spontaneous scalarization can lead to significant deviations 
of the basic neutron star properties from GR 
as demonstrated for static and slowly rotating neutron stars in
\cite{Damour:1993hw,Damour:1996ke,Harada:1998ge,Harada:1997mr,Salgado:1998sg,Sotani:2012eb,Pani:2014jra,Silva:2014fca,Sotani:2017pfj}.
Doneva and collaborators have extended these investigations
to rapidly rotating neutron stars in STT
\cite{Doneva:2013qva,Doneva:2014uma,Doneva:2014faa,Staykov:2016mbt}, 
observing that the effect of scalarization is further enhanced.
Scalarized neutron stars with a massive scalar field have also been considered
\cite{Yazadjiev:2016pcb,Doneva:2016xmf,Ramazanoglu:2016kul},
in which case the constraints on the theory are weaker, allowing, in principle, for strongly scalarized configurations with larger deviations from GR \cite{Alsing:2011er,Berti:2012bp}.

Here we investigate the effect of scalarization on static and slowly rotating
neutron stars for a large number of realistic equations of state (EOSs).
Besides a polytropic EOS, we consider two pure nuclear matter EOSs,
five EOSs describing nuclear matter with hyperons, 
four EOSs describing hybrid matter, i.e., 
nuclear matter together with quark matter,
and two EOSs for pure quark matter.
In particular, the hyperon and hybrid cases have not been considered before.
We demonstrate, that for these 14 rather different EOSs the onset of scalarization
is ruled by a single parameter of the coupling function of the scalar field. 
We then identify a strong correlation of the magnitude of the scalarization
with the metric at the center of the neutron star, independent of
the EOS. Therefore, this correlation represents an interesting
model independent result.

The search for universal relations, i.e., relations between various
physical properties of the neutron stars, which depend only a little
on the employed EOS (within certain classes of models), has been
much in the focus in recent years 
(see, e.g.,~the reviews \cite{Yagi:2016bkt,Doneva:2017}).
A basic ingredient in these relations is the compactness
${\cal C}$ of a star, which features prominently also
in the phenomenon of scalarization.
When considering the moment of inertia $I$, the tidal Love number $\lambda$
and the quadrupole moment $Q$ as functions of the compactness, one
is led to the universal $I$-Love-$Q$ relations between these quantities.

Such universal relations appear to be very valuable, 
for instance, in order to distinguish neutron stars from quark stars, 
or to test general relativity and alternative theories of gravity, 
independent of the EOS.
In STT the universal $I$-$Q$ relations \cite{Doneva:2014faa}
have been studied for rapidly rotating neutron stars.
Likewise the $I$-$\cal C$ relations \cite{Lattimer:2004nj,Breu:2016ufb}
have already been considered in STT \cite{Staykov:2016mbt},
but only for nuclear and quark matter.
Here we extend this study for our whole set of EOSs, including
the hyperon and hybrid EOS classes.

The paper is organized as follows: 
In section \ref{sec:Basic Eqs} we set up the mathematical and physical framework.
We recall the STT action,
transform from the Jordan frame to the Einstein frame,
define the scalar coupling functions,
and present the basic equations for slowly rotating neutron stars in STT.
Subsequently, we describe the set of realistic EOSs employed,
and briefly address the numerical method.
In section \ref{sec:Results} we present our results,
including the scalarized neutron star models, the analysis of the onset
and magnitude of the spontaneous scalarization, and the universal $I$-$\cal C$ relations.
We then summarize our results in section \ref{sec:Conclusions}.
Some technical details related to the analysis of the onset of 
the scalarization are given in the Appendix.
 
\section{The model}\label{sec:Basic Eqs}

\subsection{Scalar-tensor theory}\label{sec:STT}

In four dimensions, the generic action for STT (with a single scalar field)
is given in the Jordan frame by 
\cite{Damour:1992we,Damour:1996ke,Fujii:2003}
\begin{equation}
S = \frac{1}{16\pi G_{*}} \int d^4x \sqrt{-{\tilde
g}}\left[{F(\Phi)\tilde {\cal R}} - Z(\Phi){\tilde
g}^{\mu\nu}\partial_{\mu}\Phi
\partial_{\nu}\Phi   -2 U(\Phi) \right] +
S_{m}\left[\Psi_{m};{\tilde g}_{\mu\nu}\right] ,
\label{action_Jordan}
\end{equation}
where $G_{*}$ is the gravitational constant, 
$\tilde {\cal R}$ is the Ricci scalar with respect the metric ${\tilde g_{\mu\nu}}$, 
and $\Phi$ is the scalar field. 
The term $S_m$ denotes the contribution of additional matter fields to the action, 
which are parametrized into $\Psi_{m}$. 
Here we restrict to the case where the scalar field does not couple directly 
to these additional matter fields, implying that the weak equivalence principle is satisfied.
The gravitational part of the action includes the functions $F(\Phi)$ and $Z(\Phi)$,
and the potential function $U(\Phi)$. 
These functions are subject to physical restrictions,
as it was shown in 
\cite{EspositoFarese:2000ij}

For the study of neutron stars in this theory
it is convenient to change to the Einstein frame. 
This frame is related to the Jordan frame by a conformal transformation 
of the metric $g_{\mu\nu} = F(\Phi){\tilde g}_{\mu\nu}$, 
and a transformation of the scalar field 
\cite{Damour:1992we,Damour:1996ke,Fujii:2003}.
After this transformation the action becomes
\begin{equation}
S= \frac{1}{16\pi G_{*}}\int d^4x \sqrt{-g} \left[{\cal R} -
2g^{\mu\nu}\partial_{\mu}\varphi \partial_{\nu}\varphi -
4V(\varphi)\right]+ S_{m}[\Psi_{m}; \mathrm{A}^{2}(\varphi)g_{\mu\nu}] ,
\label{action_Einstein}
 \end{equation}
where ${\cal R}$ is the Ricci scalar with respect to the metric $g_{\mu\nu}$, 
and $\varphi$ is the scalar field, both being defined in the Einstein frame. 
In addition we have the following relations between the Jordan frame functions $F(\Phi)$
and $U(\Phi)$ and the Einstein frame functions $A(\varphi)$ and $V(\varphi)$
\begin{equation}
{A}(\varphi) = F^{-1/2}(\Phi) \, , \,\, 2V(\varphi) = U(\Phi)F^{-2}(\Phi) .
\end{equation}
Here we restrict to the case with vanishing scalar potential $U(\Phi)=0=V(\varphi)$. In the following we will use $c=G_{*}=1$ units unless otherwise stated.

Variation of the action (\ref{action_Einstein}) with respect to the fields
in the Einstein frame leads to the coupled set of field equations.
The Einstein equations read
\begin{equation} 
{\cal R}_{\mu\nu} - \frac{1}{2}g_{\mu\nu}{\cal R} =
  2\partial_{\mu}\varphi \partial_{\nu}\varphi   -
g_{\mu\nu}g^{\alpha\beta}\partial_{\alpha}\varphi
\partial_{\beta}\varphi
+ 8\pi T_{\mu\nu} ,
\end{equation}
where ${\cal R}_{\mu\nu}$ is the Ricci tensor, 
and $T_{\mu\nu}$ is the stress-energy tensor 
of the matter content of the action (\ref{action_Einstein}). 
The scalar field equation is given by
\begin{equation}
 \nabla^{\mu}\nabla_{\mu}\varphi = - 4\pi k(\varphi)T,
\end{equation}  
where $T = T^{\mu}_{\mu}$, 
and $k(\varphi)= \frac{d\ln({A}(\varphi))} {d\varphi}$ 
is the logarithmic derivative of the coupling function $A(\varphi)$, 
which determines the strength of the coupling 
between the scalar field and the matter.

We model the neutron star as a perfect fluid in (slow) uniform rotation. 
Hence in the physical Jordan frame 
the stress energy momentum tensor ${\tilde T}_{\mu\nu}$ is given by 
\begin{equation}
 {\tilde T}_{\mu\nu}= (\tilde \varepsilon + \tilde p){\tilde u}_{\mu}
{\tilde u}_{\nu} + {\tilde p} {\tilde g}_{\mu\nu} , 
\end{equation}
where $\tilde \varepsilon $, $\tilde p$ and $\tilde u$ denote the energy density,
the pressure and the four-velocity in the Jordan frame, respectively. 
In the Jordan frame we also assume a barotropic equation of state, 
i.e., $\tilde \varepsilon = \tilde \varepsilon(\tilde p)$.
The nuclear matter quantities $\tilde \varepsilon $, $\tilde p$ and $\tilde u$
in the Jordan frame are related to those in the Einstein frame
via the conformal factor $F(\Phi)$ 
and can be found in \cite{Damour:1992we,Damour:1996ke,Fujii:2003}.

The coupling function $A(\varphi)$ is subject to constraints from observations,
leaving however a large amount of freedom for its functional choice.
In the simple case $k(\varphi)=\kappa$, with $\kappa$ some arbitrary constant, 
a parameterization of the Brans-Dicke theory is obtained \cite{Brans:1961sx} 
where $A=e^{\kappa \varphi}$. 
Here we consider a set of two coupling functions, $A_1(\varphi)$
and $A_2(\varphi)$. The coupling function $A_1(\varphi)$ has been investigated 
widely before (see e.g.~\cite{Damour:1992we,Damour:1996ke,Doneva:2013qva,Staykov:2016mbt})
\begin{equation}
{A}_1(\varphi)=e^{\frac{1}{2}\beta\varphi^2} \ , \ \ \ k_1(\varphi)=\beta \varphi .
\label{A1}
\end{equation}
The coupling function $A_2(\varphi)$ has not yet been considered, 
and corresponds to
\begin{equation}
{A}_2(\varphi)=\frac{1}{\cosh(\sqrt{-\beta}\varphi)} \ , \ \ \
k_2(\varphi) = -\sqrt{-\beta} \tanh(\sqrt{-\beta} \varphi) . 
\label{A2}
\end{equation}
Both coupling functions have been parametrized such that they possess the
same quadratic expansion coefficient. They differ only in higher order,
where the fall-off of ${A}_2(\varphi)$ is slower.
This is in contrast to the coupling function $A_3(\varphi) = \cos{(\sqrt{-\beta}\varphi)}$
employed in \cite{Damour:1993hw}, which exhibits a faster fall-off.
Note that all of these three couplings are invariant 
under the transformation $\varphi \to -\varphi$. 

The strongest observational constraint on the possible values of the constant $\beta$ 
that should be taken into account comes 
from the binary pulsar PSR J1738+0333 \cite{Freire:2012mg},
which requires 
\begin{equation}
\frac{d^{2} \ln({\cal  A}(\varphi))}{d\varphi^{2}}|_{\varphi=0}=\beta \ge -4.5 .
\end{equation}

\subsection{Slowly rotating neutron stars in scalar-tensor theory}

In order to describe slowly rotating neutron stars, 
we choose the following form of the metric in the Einstein frame
\begin{equation}
ds^{2}=-e^{f(r)}dt^{2}+\frac{1}{n(r)}d r^{2} + r^{2} d \theta^{2} 
+r^{2}\sin^{2}\theta(d \phi+ \xi \omega(r)dt)^{2} ,
\label{eq:metric}
\end{equation}
where the metric functions $f(r)$, $n(r)$ and $\omega(r)$ 
depend only on the radial coordinate $r$. 
We introduce $\xi$ as a perturbation theory parameter,
that allows us to keep track of the slow rotation approximation,
i.e., all expressions are to be considered up to $O(\xi^2)$. 

The inertial dragging $\omega(r)$ vanishes in the static case. 
In the slow rotation approximation
the scalar field is not affected by the rotation,
since $\varphi= \varphi(r) + O(\xi^2)$, 
and hence it is only a function of the radial coordinate. 
The same applies to the energy density 
$\tilde \varepsilon= \tilde \varepsilon(r) + O(\xi^2)$ 
and the pressure $\tilde p=\tilde p(r) + O(\xi^2)$. 
The four velocity of the fluid in the slow rotation approximation 
is $\tilde u = u^t (\partial_t + \xi \Omega\partial_{\phi})$, 
where $\Omega$ is the angular velocity of the fluid. 

With the metric ansatz Eq.~(\ref{eq:metric}) 
and the above definitions
the Einstein field equations in the slow rotation approximation
reduce to
a
system of Ordinary Differential Equations (ODEs)
that has been presented before in the literature
\cite{Damour:1993hw,Damour:1996ke,Harada:1998ge,Salgado:1998sg,Sotani:2012eb}
.

Regularity of the configurations at the center of the star ($r=0$) 
imposes 
a particular expansion in terms of the radial coordinate $r$,
which can be found in \cite{Pani:2014jra}.

The surface of the star is defined as the surface of constant radius $r=R$,
where the pressure vanishes, $\tilde p|_{R}=0$. 
The exterior of the star is then given by $r>R$. 
Here the energy density and the pressure vanish: 
$\tilde p |_{r>R}=\tilde \varepsilon |_{r>R}=0$. 
However, the scalar field does not vanish outside the star,
when the star is scalarized. 
Note that the physical radius of the star is defined in the Jordan frame,
i.e., $R_s = R A(\varphi(R))$.

Since we require the solutions to be asymptotically flat, 
the functions exhibit the following behaviour close to infinity \cite{Damour:1993hw,Harada:1998ge,Salgado:1998sg,Sotani:2012eb,Pani:2014jra,Silva:2014fca}
\begin{equation}
 m(r) = M - \frac{1}{2}\frac{\omega_A^2}{r} - \frac{1}{2}\frac{\omega_A^2 M}{r^2} 
+O(\frac{1}{r^3}) ,
\end{equation}
\begin{equation}
 f(r) = -\frac{2M}{r} - \frac{2M^2}{r^2} - \frac{1}{3}\frac{M(M^2-\omega_A^2)}{r^3} 
+O(\frac{1}{r^4}) ,
\end{equation}
\begin{equation}
 \varphi(r) = \frac{\omega_A}{r} + \frac{M\omega_A}{r^2} 
 +\frac{1}{6}\frac{\omega_A(8M^2-\omega_A^2)}{r^3}
+O(\frac{1}{r^3}) ,
\end{equation}
\begin{equation}
\omega(r) = 
\frac{2J}{r^3}
+O(\frac{1}{r^5}) ,
\end{equation}
where we have defined the function 
$m(r)=(1-n(r))r/2$. Note that here we restrict to the case $\varphi\rvert_{\infty}=0$.

From the asymptotic behaviour of the functions 
we can extract a number of physical properties of the stars. 
For instance, provided that $\varphi\rvert_{\infty}=0$, 
the physical mass of the star is simply given by $M$, 
and the angular momentum of the star is given by $J$. 
The moment of inertia $I$ is then calculated as the ratio of 
the angular momentum and the angular velocity of the fluid
\begin{equation}
I = J/\Omega .
\end{equation}
In addition, if the scalar field is nontrivial, 
the neutron star possesses scalar hair, characterized by the scalar charge $\omega_A$.

Although the expansion at the origin and the asymptotic expansion 
depend on a number of undetermined parameters, 
a full solution of the set of coupled equations depends on fewer parameters. 
Indeed, once the equation of state is provided 
($\tilde \varepsilon=\tilde \varepsilon(\tilde p)$)
and the coupling function $A(\varphi)$ is fixed, 
a solution depends only on two parameters, 
the mass $M$ and the angular momentum $J$.
(In first order perturbation theory the angular momentum $J$
is proportional to the angular velocity $\Omega$.) 
The scalar charge $\omega_A$, if present, 
is only a function of the mass, 
and hence can be considered as to represent only secondary scalar hair.

\subsection{Equations of State}\label{sec:EOS}

As commented on above, in order to integrate the system of equations
we have to provide an equation of state 
in the form $\tilde \varepsilon=\tilde \varepsilon(\tilde p)$. 
Here we consider a large number of realistic EOSs, 
obtained from effective models of the nuclear interactions 
subject to different assumptions.

In order to compare the effects of exotic matter 
in the properties of the configurations, 
we have studied two EOSs containing 
only nuclear matter: 
SLy \cite{Douchin:2001sv} and APR4 \cite{Akmal:1998cf}. 
For EOSs containing nucleons and hyperons 
we have considered the following five cases: 
BHZBM \cite{Bednarek:2011gd}, GNH3 \cite{Glendenning:1984jr}, 
H4 \cite{Lackey:2005tk} and WCS1, WSC2 \cite{Weissenborn:2011ut}. 
For pure quark matter we use two EOSs: 
WSPHS1 and WSPHS2 \cite{Weissenborn:2011qu}. 
For hybrid matter consisting of quarks and nucleons 
we consider these four EOSs: 
ALF2, ALF4 \cite{Alford:2004pf}, BS4 \cite{Bonanno:2011ch} 
and WSPHS3 \cite{Weissenborn:2011qu}.

In addition, for completeness, we also include 
the results for a polytropic EOS
\begin{equation}
\tilde \varepsilon = 
K \frac{{\tilde \rho}^\Gamma}{\Gamma - 1} + {\tilde \rho} \ , \ \ \
\tilde p=  K {\tilde \rho}^\Gamma \ , \ \ \
\Gamma =  1 + \frac{1}{N} ,
\label{poly}
\end{equation}
where $\tilde \rho$ is the baryonic mass density, 
and we have chosen for the polytropic constant $K =1186.0$, 
and for the adiabatic index $\Gamma$ the polytropic index $N=0.7463$. 

All the EOSs considered possess 
a maximum mass close to or larger than $2 M_{\odot}$, 
which is the current maximum mass observed 
in neutron star candidates (PSR J1614-2230 \cite{Demorest:2010bx} 
and PSR J0348+0432 \cite{Antoniadis:2013pzd}).

\subsection{Numerical Method}\label{sec:Numerical Method}

The configurations of slowly rotating neutron stars 
are generated numerically by solving the stellar structure equations 
with appropriate boundary conditions which ensure regularity at the center 
and asymptotic flatness.

For the numerical integration of this coupled set of ODEs,
we use the ODE solver package COLSYS \cite{Ascher:1979iha}. 
This code allows to numerically solve boundary value problems 
for systems of nonlinear coupled ODEs, and is equipped 
with an adaptive mesh selection procedure. 
 
The solution is required to be regular at the center of the star, 
and to approach at infinity the Minkowski metric, 
with the scalar field vanishing there \cite{Damour:1993hw,Damour:1996ke,Harada:1998ge,Salgado:1998sg,Sotani:2012eb}.

For the numerical integration, it is useful to compactify 
space by a transformation of the radial coordinate 
\begin{equation}
y(r) = \frac{r}{r+R} ,
\end{equation}
where $r=R$ determines the surface of the star,
i.e., the surface of the star resides at $y=1/2$. 
We integrate the resulting set of equations 
in the region $y\in [0 , 1]$.
In order to compute the coordinate radius $R$, 
we introduce an auxiliary differential equation, 
\begin{equation}
\frac{dR}{dy}=0 . 
\end{equation}
The system of ODEs is then complemented with a 
further boundary condition at the surface of the star, 
\begin{equation}
\tilde p|_{\frac{1}{2}}=0 .
\end{equation}

The EOSs are implemented using different methods. 
The case of the relativistic polytrope is the simplest one, 
since the relation 
$\tilde \varepsilon=\tilde \varepsilon(\tilde p)$ is known analytically. 

The EOSs corresponding 
to WCS1, WCS2, WSPHS1, WSPHS2, WSPHS3, BS4 and BHZBM 
are available in table form. 
Hence for these cases we use a piecewise monotonic 
cubic Hermite interpolation of the data points.

For the equations SLy, APR4, GNH3, H4 and ALF2, ALF4 
we implement in the code the piecewise polytropic interpolation 
presented in \cite{Read:2008iy}. 
In this interpolation different regions of the EOS
are approximated as specific polytropes.

\section{Results} \label{sec:Results}

In this section we present our results 
for static and slowly rotating neutron stars for 14 realistic EOSs in STT,
employing the two coupling functions
$A_1=e^{\frac{1}{2}\beta\varphi^2}$ and $A_2=1/\cosh(\sqrt{-\beta}\varphi)$. 
In particular, we present results for $\beta_1=-4.8$, 
which already violates the constraint obtained 
from pulsar observations \cite{Freire:2012mg}, 
and $\beta_2=-4.5$, which is currently the 
largest negative value of $\beta$ allowed by observations.

We note, that the GR configurations are also solutions 
of the full scalar tensor theory, since in the case $\varphi=0$, 
the equations reduce to Einstein gravity.

\subsection{Neutron star models}  \label{sec:Models}

\begin{figure}[p!]
     \centering
\includegraphics[angle=-90,width=0.85\linewidth]{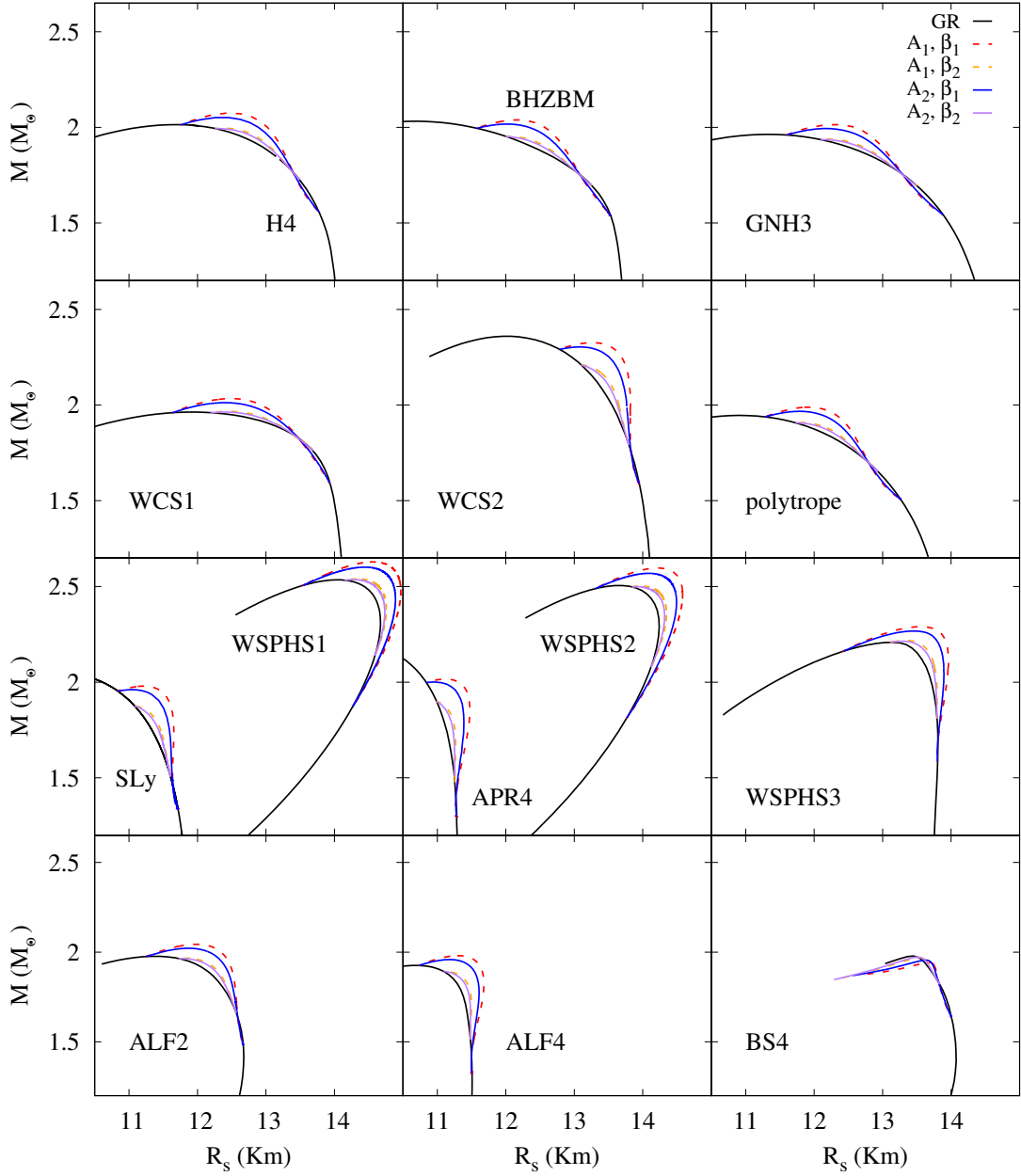}
     \caption{
Total mass $M$ (in solar masses $M_\odot$) 
versus the physical radius $R_s$ (in km) of the neutron star models 
for all EOSs considered:
The first two rows show the 5 hyperon EOSs 
(H4, BHZBM, GNH3, WCS1, WCS2)
and the polytropic EOS,
the last two rows contain the 2 nuclear EOSs (SLy, APR4),
the 2 quark EOSs (WSPHS1, WSPHS2) and the 4 hybrid EOSs
(WSPHS3, ALF2, ALF4, BS4).
The solid black lines represent the GR configurations. 
The dashed red and orange lines represent the scalarized solutions 
for $A_1=e^{\frac{1}{2}\beta\varphi^2}$ 
with $\beta_1=-4.8$ and $\beta_2=-4.5$, respectively. 
The solid blue and purple lines represent the scalarized solutions 
for $A_2=1/\cosh(\sqrt{-\beta}\varphi)$ 
with the same values of $\beta_1=-4.8$ and $\beta_2=-4.5$.
}
     \label{plot_MR_12}
\end{figure}

\begin{figure}[p!]
     \centering
\includegraphics[angle=-90,width=0.85\linewidth]{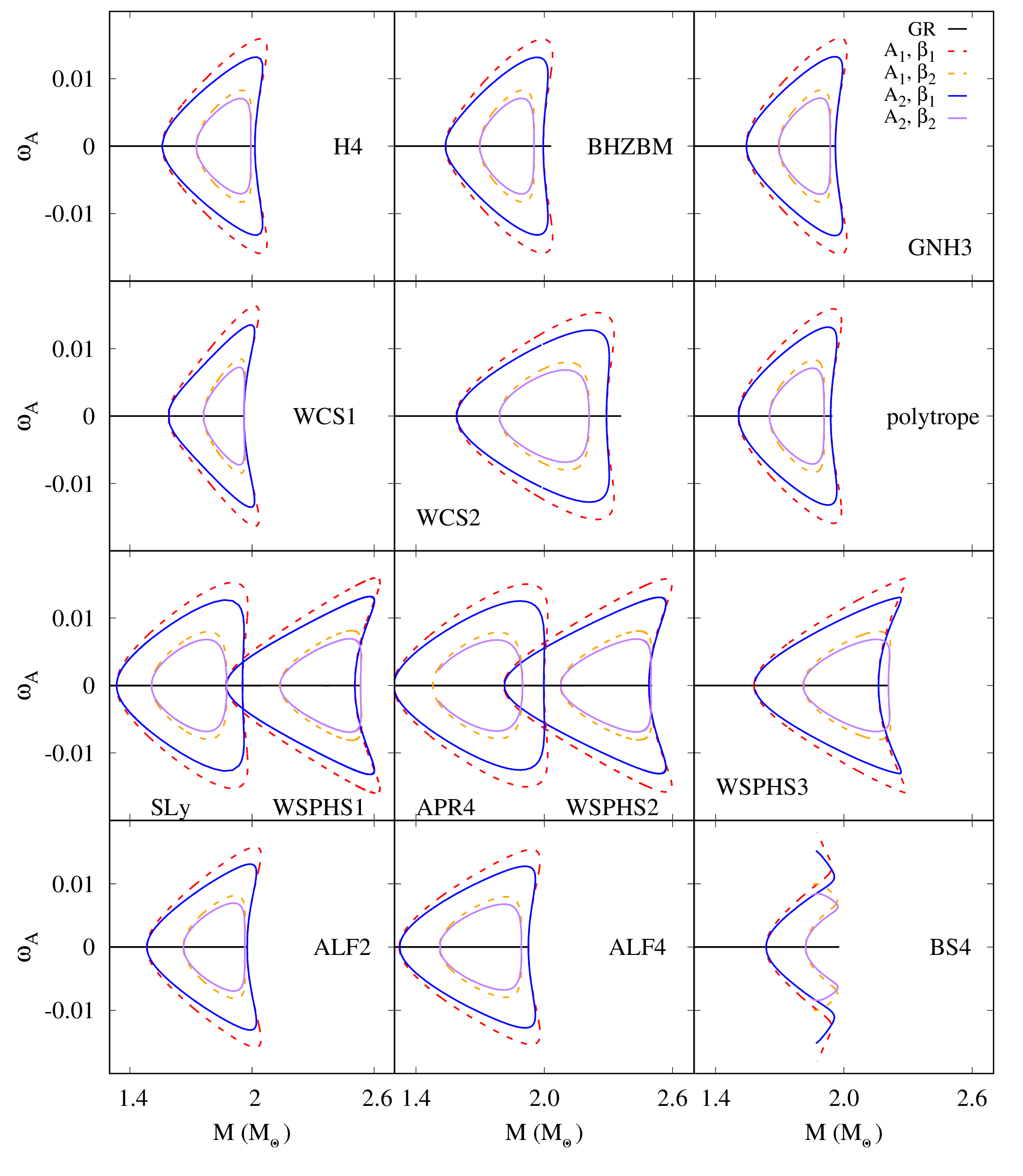}
     \caption{
Scalar field charge $\omega_A$ versus the
total mass $M$ (in solar masses $M_\odot$) of the neutron star models
for all EOSs considered:
The first two rows show the 5 hyperon EOSs
(H4, BHZBM, GNH3, WCS1, WCS2)
and the polytropic EOS,
the last two rows contain the 2 nuclear EOSs (SLy, APR4),
the 2 quark EOSs (WSPHS1, WSPHS2) and the 4 hybrid EOSs
(WSPHS3, ALF2, ALF4, BS4).
The solid black lines represent the GR configurations.
The dashed red and orange lines represent the scalarized solutions
for $A_1=e^{\frac{1}{2}\beta\varphi^2}$
with $\beta_1=-4.8$ and $\beta_2=-4.5$, respectively.
The solid blue and purple lines represent the scalarized solutions
for $A_2=1/\cosh(\sqrt{-\beta}\varphi)$
with the same values of $\beta_1=-4.8$ and $\beta_2=-4.5$.
Note that the scalar charge can be positive and negative.}
     \label{plot_omAM_12}
\end{figure}

We now present our results for the static neutron star models,
showing the total mass $M$ (in solar masses $M_\odot$)
versus the physical radius $R_s$ (in km) 
in Fig.~\ref{plot_MR_12},
and the scalar field charge $\omega_A$ versus the
total mass $M$ (in solar masses $M_\odot$)
in Fig.~\ref{plot_omAM_12}.

In these two figures all 14 EOSs are considered in the same succession.
The first two rows show the 5 EOSs containing hyperons and nucleons
(H4, BHZBM, GNH3, WCS1, WCS2)
and the polytropic EOS,
the last two rows contain the 2 nuclear EOSs (SLy, APR4),
the 2 quark EOSs (WSPHS1, WSPHS2), where we have superimposed
(SLy,  WSPHS1) and (APR4, WSPHS2), as well as the 4 hybrid EOSs
containing quarks and nucleons
(WSPHS3, ALF2, ALF4, BS4).

The scalarized neutron star models have been computed
for the scalar coupling $A_1$
for the coupling constants $\beta_1=-4.8$ (dashed red)
and $\beta_2=-4.5$ (dashed orange),
and the scalar coupling $A_2$
for the same coupling constants $\beta_1=-4.8$ (solid blue)
and $\beta_2=-4.5$ (solid purple).
The GR configurations are always included as well
(solid black).

The mass--radius curves in Fig.~\ref{plot_MR_12} show
a number of interesting facts.
The onset of scalarization depends only on the value of $\beta$,
i.e., it is the same for the coupling functions $A_1$ and $A_2$,
and it would also be the same for the coupling function $A_3$
\cite{Damour:1993hw,Damour:1996ke}.
Thus it is determined only by the coefficient 
of the quadratic term in $\varphi$,
and the lower the value of $\beta$
the stronger is the effect of scalarization.
Since the coupling functions differ in their higher order terms,
and $A_1$ decreases faster than $A_2$, the scalarization
is stronger for $A_1$ than for $A_2$.
Likewise, it is stronger for $A_3$ than for $A_1$
\cite{Damour:1993hw,Damour:1996ke}.

From Fig.~\ref{plot_MR_12} we see
that for the observational limit $\beta_2$ 
one obtains typically scalarized solutions 
with masses below the maximum GR mass. 
The exceptions are WSPHS1, WSPHS2, and WSPHS3
(quark matter and hybrid matter),
where the maximum mass of the scalarized configurations 
is slightly larger than the GR maximum mass. 

Note that the hybrid EOS BS4 is a very special case.
In particular, the EOS table for BS4 we have employed 
does not contain values for sufficiently high densities,
i.e., values where the scalarization is expected to vanish again.
Therefore the scalarized branches here simply stop
without being able to merge again with GR solutions,
when the scalarization vanishes again.

Let us note
that the onset of the scalarization
is not strongly correlated with the
value of the central density.
While the central density at the onset 
is of the same order of magnitude in all cases, 
it can differ by a factor of two for the different EOSs.
The same is true, when the onset of
scalarization is considered versus the central pressure.
Therefore, both quantities are not good indicators
of scalarization.
The trace $T$ of the energy-momentum tensor is even
worse. Here even the sign of $T$ can differ
for different EOSs.

Still, as seen in Fig.~\ref{plot_omAM_12}, 
where the scalar field charge $\omega_A$ 
is exhibited as a function of the total mass $M$
of the stars, 
the general behaviour and the maximum value
of the scalar charge are very similar for all EOSs --
except for BS4 (where the results suggest that beyond the maximum mass, the scalarized configurations cannot be trusted). 
This is surprising since the EOSs describe
physically widely differing systems,
and it calls for further investigation to be performed
in the next subsection.
We note that the symmetry $\varphi \to -\varphi$
of the equations implies the symmetry $\omega_A \to -\omega_A$.

\subsection{Onset and magnitude of the scalarization} \label{sec:Onset}

\begin{figure}[t!]
\begin{center}
\mbox{\hspace{0.2cm}
\includegraphics[height=.45\textheight, angle =-90]{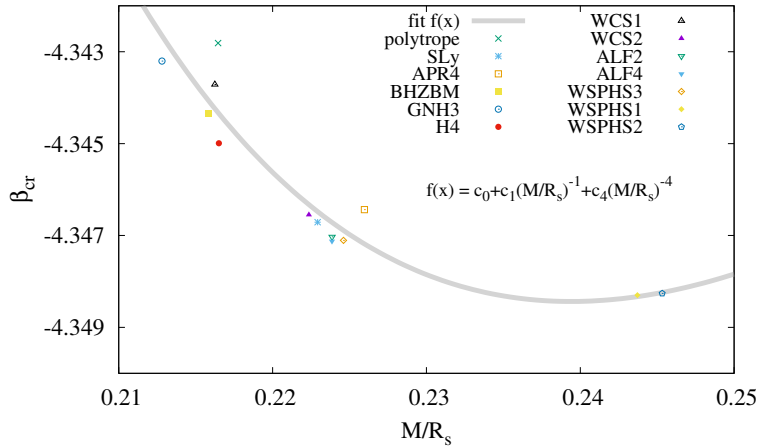}
}
\end{center}
\vspace{-0.5cm}
\caption{The critical value of the coupling parameter $\beta$ versus 
the compactness ${\cal C}=M/R_s$ for all EOSs considered (except BS4). 
The grey curve is a fit to the function $f=c_0 + c_1 (M/R)^{-1} + c_4 (M/R)^{-4}$.
}
\label{Fig_beta_cr_MR}
\end{figure}

As noted above,
since in the limit of small scalar field $\varphi$, 
the coupling functions considered are essentially the same 
($A_1 \sim A_2 \sim 1 + \frac{1}{2}\beta\varphi^2 + ...$), 
the branching configurations, where scalarization begins 
and ends, coincide for a given EOS
for the coupling functions $A_i(\varphi)$, 
when the coupling parameter $\beta$ has the same value.
This holds, in particular, also for the
critical values $\beta_{\rm cr}$,
which determine the onset of scalarization.

In Fig.~\ref{Fig_beta_cr_MR} $\beta_{\rm cr}$ is shown
versus the compactness ${\cal C}=M/R_s$ for all EOSs considered,
except for BS4. For BS4 the onset arises at $\beta_{\rm cr}=-4.336$
and ${\cal C}=M/R_s=0.2193$, and thus differs $\% 0.2$ in $\beta$.  
For all other EOSs, 
thus including nuclear, hyperon, hybrid and quark matter,
the value of $\beta_{\rm cr}$ varies only between
-4.348 and -4.343,
in good agreement with the value of -4.35 given by Harada
\cite{Harada:1998ge}.
It was also noted by Harada \cite{Harada:1997mr}
that there exists a relation between the region of scalarization and the compactness of the star.
As seen in Fig.~\ref{Fig_beta_cr_MR},
the onset of scalarization can be well parametrized by
the function of the compactness
\begin{equation}
f=c_0 + c_1 (M/R)^{-1} + c_4 (M/R)^{-4} 
\end{equation}
with $c_0=-4.17789$, $c_1=-0.0544455$ and $c_4 = 0.000186767$,
and the reduced $\chi^{2} $ is $4.3 \cdot 10^{-7}$.
An efficient method to obtain $\beta_{\rm cr}$
is discussed in the Appendix.

\begin{figure}
\hspace*{-0.5cm}
 \begin{subfigure}[b]{0.45\textwidth}
\includegraphics[width=72mm,scale=1,angle=-90]{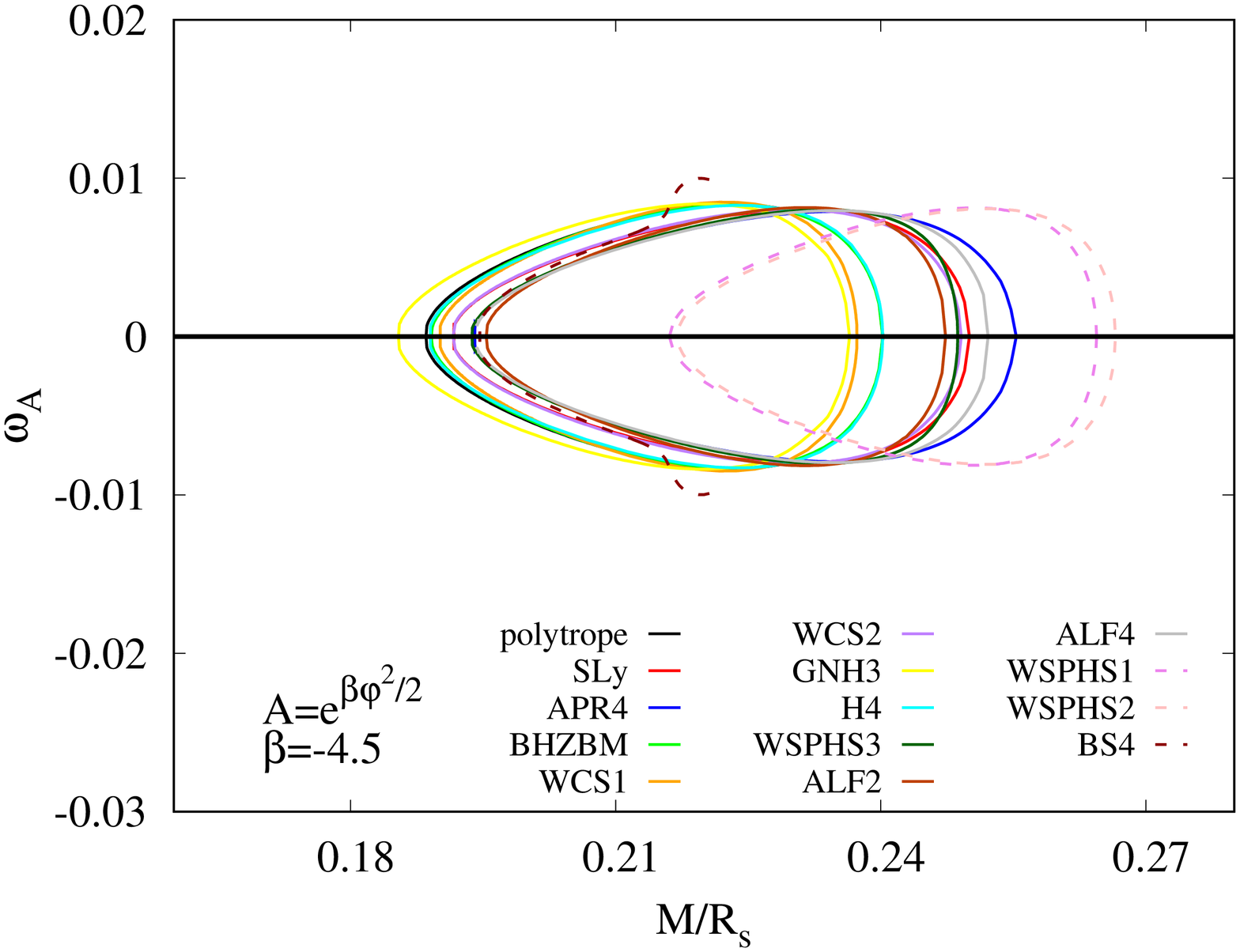}
         \caption{}
         \label{plot_omA_MR_14_beta:a}
     \end{subfigure}
\hspace*{0.8cm}
     \begin{subfigure}[b]{0.45\textwidth}
\includegraphics[width=72mm,scale=1,angle=-90]{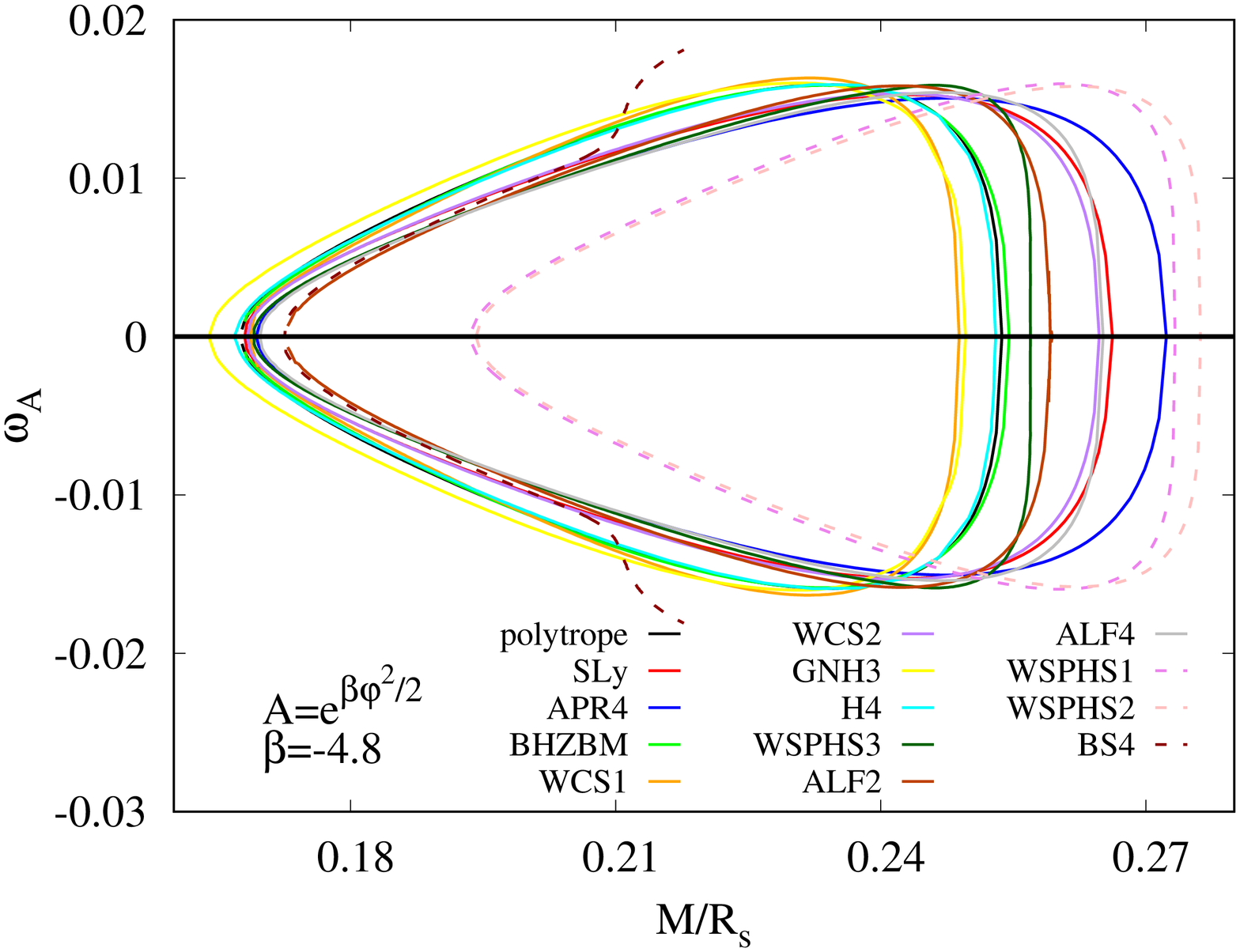}
         \caption{}
         \label{plot_omA_MR_14_beta:b}
     \end{subfigure}
\caption{
Scalar field charge $\omega_A$ versus the
compactness ${\cal C}=M/R_s$ (in units of $c=G_{*}=1$)
of the neutron star models
for all EOSs considered:
The 5 hyperon EOSs
(H4, BHZBM, GNH3, WCS1, WCS2),
the polytropic EOS,
the 2 nuclear EOSs (SLy, APR4),
the 2 quark EOSs (WSPHS1, WSPHS2) and the 4 hybrid EOSs
(WSPHS3, ALF2, ALF4, BS4).
The coupling function is $A_1=e^{\frac{1}{2}\beta\varphi^2}$
with $\beta_2=-4.5$ in (a) and  $\beta_1=-4.8$ in (b).
Note that the scalar charge can be positive and negative.}
     \label{plot_omA_MR_14_beta}
\end{figure}

\begin{figure}
\begin{subfigure}[b]{0.45\textwidth}
\hspace{-0.5cm}
\includegraphics[width=72mm,scale=1,angle=-90]{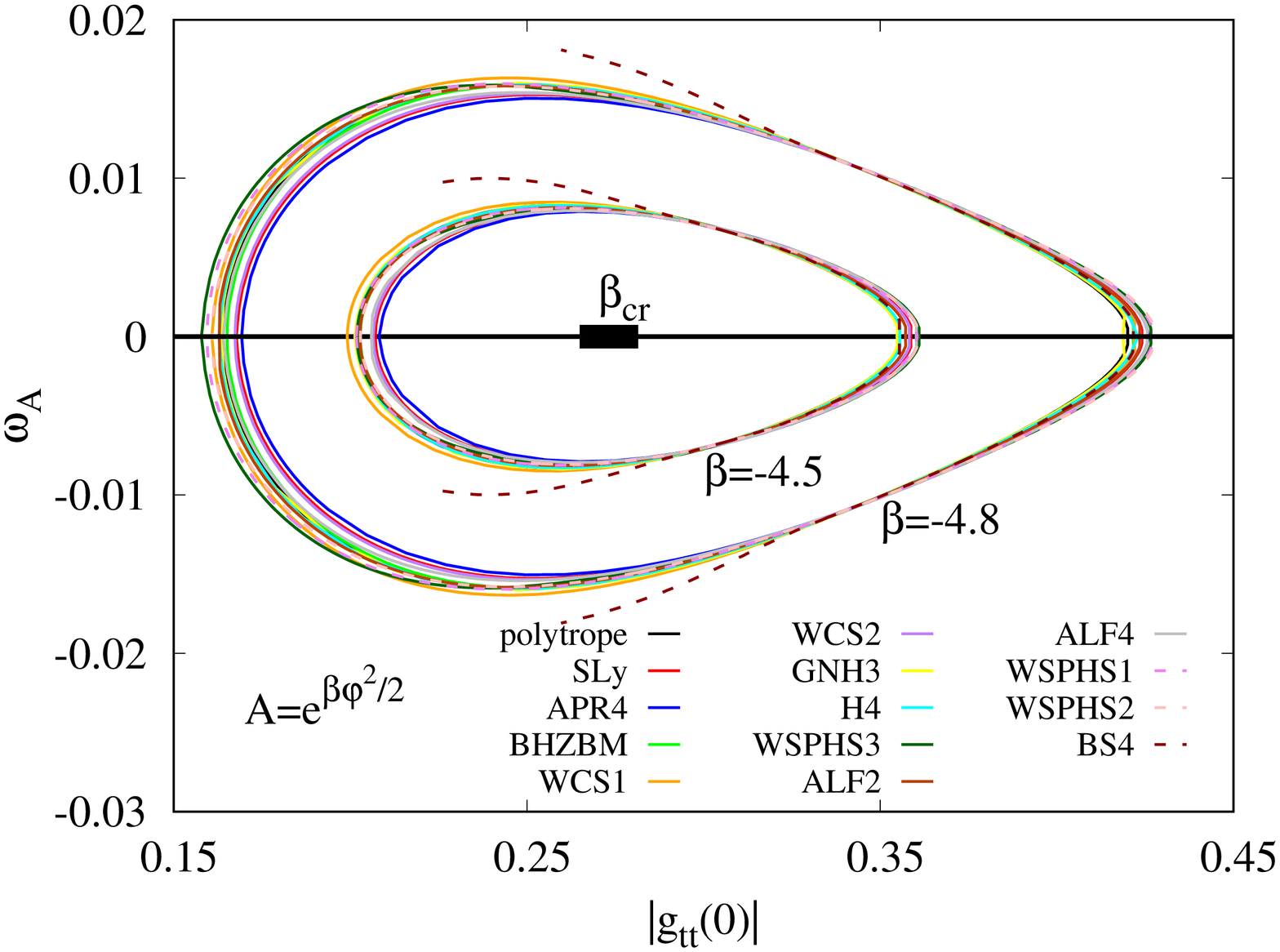}
\caption{}
         \label{plot_omA_gtt_14:a}
     \end{subfigure}
\begin{subfigure}[b]{0.45\textwidth}
\hspace{0.8cm}
\includegraphics[width=72mm,scale=1,angle=-90]{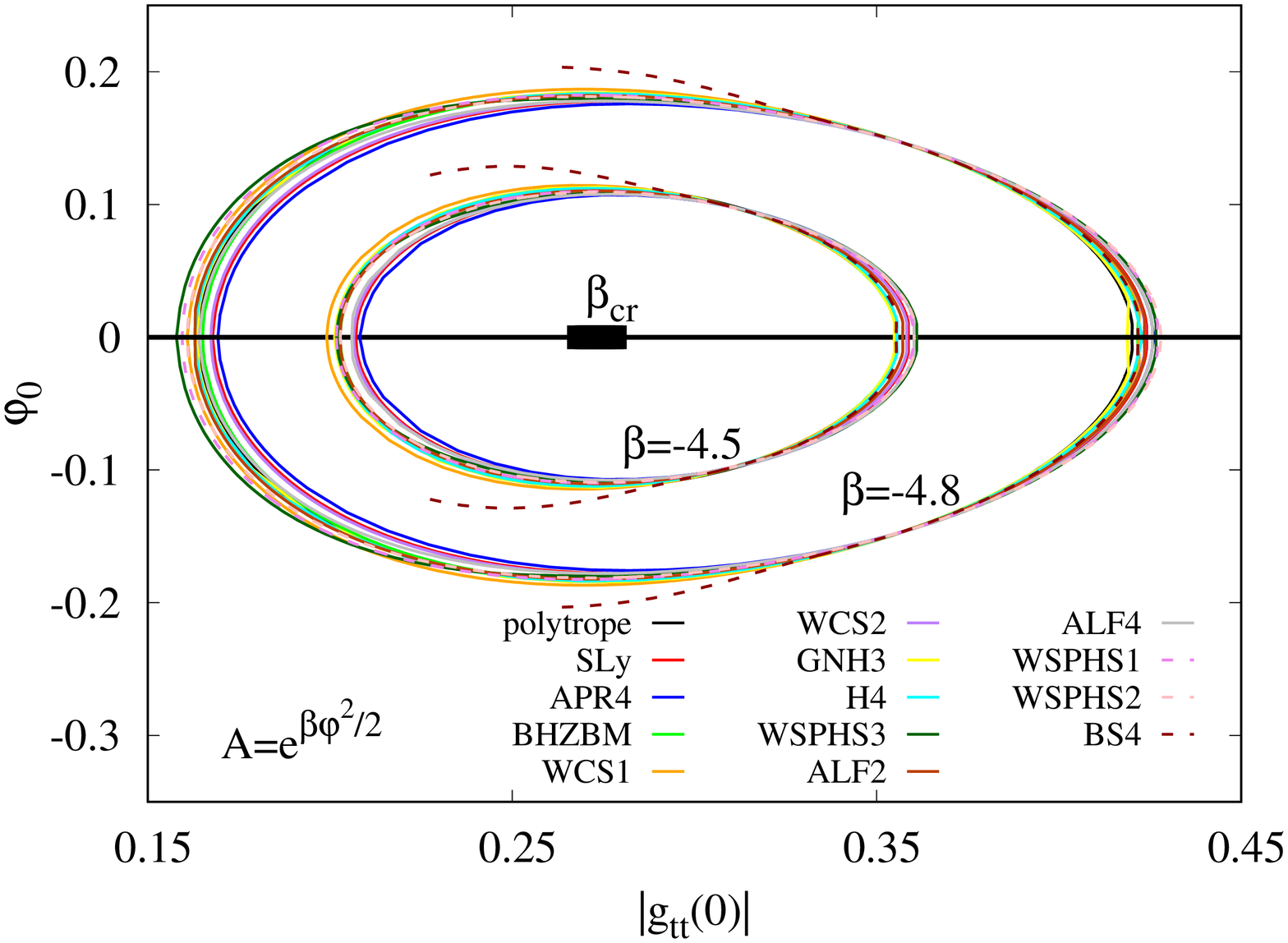}
\caption{}
         \label{plot_omA_gtt_14:b}
     \end{subfigure}
\caption{
Scalar field charge $\omega_A$ (a) and scalar field at the center
$\varphi_0$ (b) versus the
metric function $g_{tt}(0)$ at the center of the neutron star models
for all EOSs considered:
The 5 hyperon EOSs
(H4, BHZBM, GNH3, WCS1, WCS2),
the polytropic EOS,
the 2 nuclear EOSs (SLy, APR4),
the 2 quark EOSs (WSPHS1, WSPHS2) and the 4 hybrid EOSs
(WSPHS3, ALF2, ALF4, BS4).
The coupling function is $A_1=e^{\frac{1}{2}\beta\varphi^2}$
with $\beta_1=-4.8$ and $\beta_2=-4.5$, respectively.
The critical values $\beta_{\rm cr}$ are also indicated.
Note that the scalar charge and the scalar field can be positive and negative.}
     \label{plot_omA_gtt_14}
\end{figure}

\begin{table}
\centering
\begin{tabular}{||c||c|c|c|c||}
    \cline{2-5}
    \multicolumn{1}{c||}{} 
 & {$\boldsymbol{A_1}$, $\boldsymbol{\beta_1}$} & {$\boldsymbol{A_1}$, $\boldsymbol{\beta_2}$} & {$\boldsymbol{A_2}$, $\boldsymbol{\beta_1}$} & {$\boldsymbol{A_2}$, $\boldsymbol{\beta_2}$} \\ 
\hline\hline 
\textbf{$\boldsymbol{E[|\omega_A(max)|]}$}        & $0.0157$ & $0.00815$ & $0.0130$ & $0.00696$ \\ 
\hline 
\textbf{$\boldsymbol{CV}$} & $2.28\cdot 10^{-2}$ & $2.29\cdot 10^{-2}$  & $2.11\cdot 10^{-2}$ & $2.21\cdot 10^{-2}$ \\ 
\hline 
\textbf{$\boldsymbol{E[|\varphi_0(max)|]}$}        & $0.18138$ & $0.11040$ & $0.15810$ & $0.09593$ \\ 
\hline 
\textbf{$\boldsymbol{CV}$} & $1.74\cdot 10^{-2}$ & $2.14\cdot 10^{-2}$ & $1.75\cdot 10^{-2}$ & $2.12\cdot 10^{-2}$ \\ 
\hline 
\textbf{$\boldsymbol{E[|\varphi_s(max)|]}$}        & $0.10000$ & $0.05864$ & $0.08620$ & $0.05076$ \\ 
\hline 
\textbf{$\boldsymbol{CV}$} & $5.60\cdot 10^{-2}$ & $5.86\cdot 10^{-2}$ & $5.81\cdot 10^{-2}$ & $5.94\cdot 10^{-2}$ \\ 
\hline 
\end{tabular} 
\caption{Mean value and coefficient of variation
for the maximum value
of the scalar field charge $\omega_A$, 
the central value of the scalar field $\varphi_0$, 
and the surface value of the scalar field $\varphi_s$, 
considering all EOSs except BS4.}
\label{tab:scalarECV}
\end{table}

As seen in Fig.~\ref{plot_omAM_12}, the general behaviour of the 
scalar field is quite independent of the EOS considered,
with the exception of the EOS BS4. 
Let us therefore now consider the
mean values $E$ and the corresponding coefficients of variation ($CV$)
(i.e., the ratio of the standard deviation to the mean value)
of several characteristic properties of the scalar field,
obtained for the full set of EOSs except for BS4.
In particular, we exhibit in Table \ref{tab:scalarECV}
for both coupling functions $A_1$ and $A_2$
and both coupling parameters $\beta_1$ and $\beta_2$
the mean value and the coefficient  of variation
for the maximum value of the scalar field charge $\omega_A$,
the central value of the scalar field $\varphi_0$,
and the surface value of the scalar field $\varphi_s$.
Clearly, the $CV$ is rather small for all quantities.

The above analysis indicates again, 
that there should be some largely EOS independent agent 
responsible for the magnitude of the scalarization.
Let us then again consider the compactness $\cal C$
and study the dependence of the scalar field
on the compactness, which is the major ingredient
for many universal relations.
To that end we exhibit in Fig.~\ref{plot_omA_MR_14_beta}
the scalar field charge $\omega_A$ versus the
the compactness ${\cal C}=M/R_s$ 
of all neutron star models for all 14 EOSs
with coupling function $A_1$
for $\beta=-4.5$ [Fig.~\ref{plot_omA_MR_14_beta}(a)]
and $\beta=-4.8$ [Fig.~\ref{plot_omA_MR_14_beta}(b)].

These figures indeed reveal a certain amount of EOS
independence, showing some clustering in the small $\cal C$
region for the nuclear, hyperon and hybrid EOSs.
However, the two quark EOSs are distinctly offset,
being shifted to higher compactness.
The figures also show that the BS4 EOS follows the
general trend of the nuclear, hyperon and hybrid EOSs
only up to a certain compactness (close to the maximum value of the mass), where it starts
to behave strangely.
This possibly indicates that 
beyond this certain compactness this EOS
may no longer be reliable in this context.

While compactness is certainly an important ingredient,
it does not fully predict the onset and magnitude
of the scalarization.
A much better predictor of the onset and magnitude of the
scalarization is the gravitational potential at the
center of the star as embodied by the metric component $g_{tt}(0)=-e^{f(0)}$.
Note this expression is not coordinate dependent, since the gauge freedom has
been fixed by specifying the metric in Eq. (\ref{eq:metric}).
We have investigated the dependence of the scalar field charge $\omega_A$,
and of the values of the scalar field at the center $\varphi(0)$
and at the surface versus the value of the
metric function $g_{tt}(0)$ at the center of the star 
for all EOSs and both of the coupling functions. 
This dependence is demonstrated in Fig.~\ref{plot_omA_gtt_14},
where we show the results for the coupling function $A_1$. 
Indeed, there is a strong universal behaviour visible,
including all EOSs, also the quark EOSs. 
Only the BS4 EOS starts to deviate again, and should
possibly no longer be trusted beyond the maximum mass.

\subsection{Universal $I$--$\cal C$ relations} \label{sec:Universal}

\begin{figure}
     \centering
\includegraphics[angle=-90,width=0.85\linewidth]{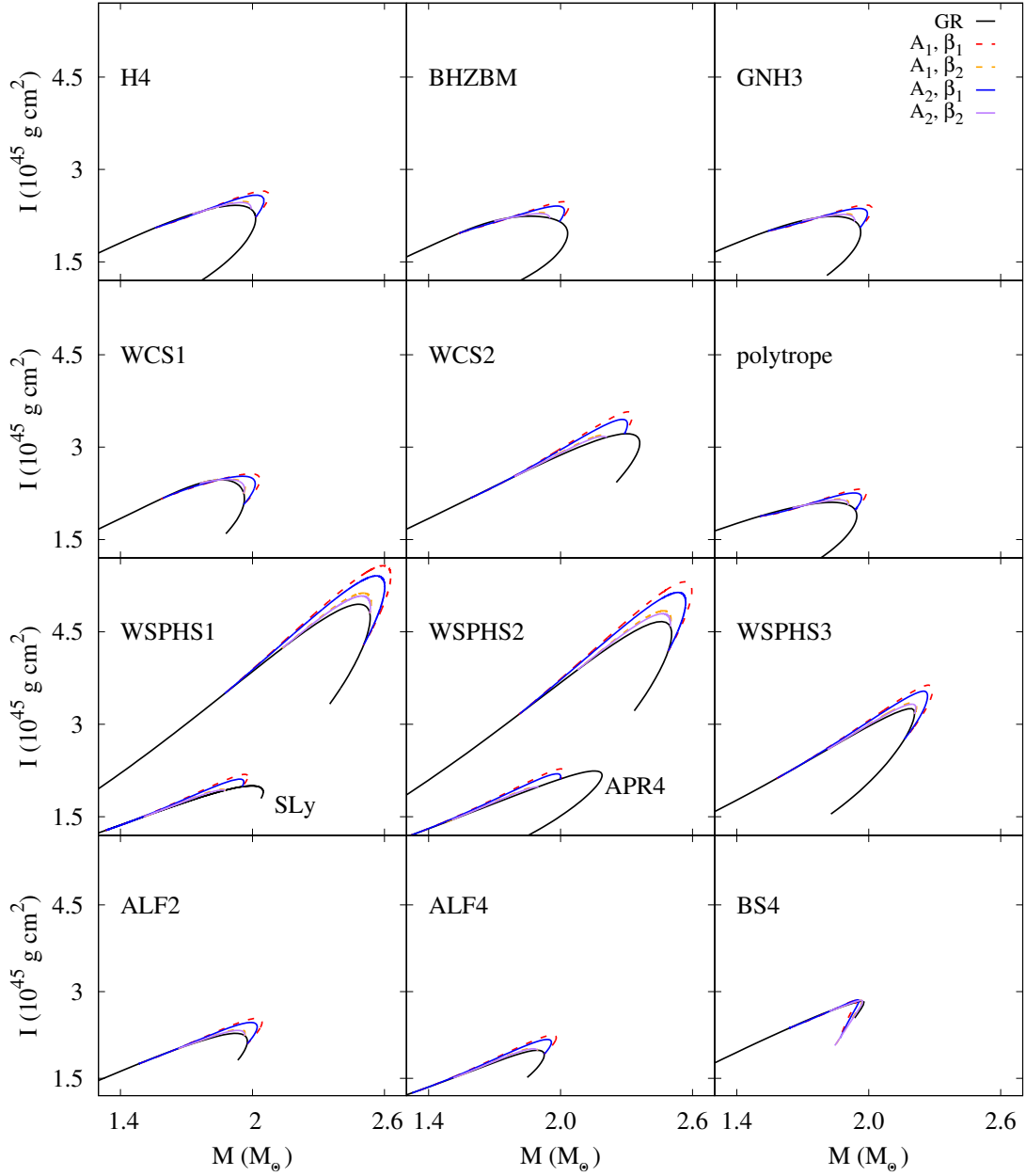}
     \caption{
Moment of inertia $I$ (in $10^{45}$ g cm$^2$) versus 
the total mass $M$ (in solar masses $M_\odot$)
of the neutron star models for all EOSs considered:
The first two rows show the 5 hyperon EOSs
(H4, BHZBM, GNH3, WCS1, WCS2)
and the polytropic EOS,
the last two rows contain the 2 nuclear EOSs (SLy, APR4),
the 2 quark EOSs (WSPHS1, WSPHS2) and the 4 hybrid EOSs
(WSPHS3, ALF2, ALF4, BS4).
The solid black lines represent the GR configurations.
The dashed red and orange lines represent the scalarized solutions
for $A_1=e^{\frac{1}{2}\beta\varphi^2}$
with $\beta_1=-4.8$ and $\beta_2=-4.5$, respectively.
The solid blue and purple lines represent the scalarized solutions
for $A_2=1/\cosh(\sqrt{-\beta}\varphi)$
with the same values of $\beta_1=-4.8$ and $\beta_2=-4.5$.
}
     \label{plot_IM_12}
\end{figure}

We now turn to slowly rotating neutron star models, obtained
in lowest order perturbation theory.
In Fig.~\ref{plot_IM_12} we present the moment of inertia $I$ 
as a function of the total mass $M$ of the neutron stars. 
The moment of inertia represents an important physical property
of the neutron stars, since it can be obtained from timing observations
of pulsars, and thus represents another observational handle
to constrain the EOS of neutron stars.
As seen in Fig.~\ref{plot_IM_12}
the effect of the scalarization is to allow for somewhat larger values
of the moment of inertia than in GR. 

Let us now address the universality of the moment-of-inertia--compactness
relations, suggested before \cite{Lattimer:2004nj,Breu:2016ufb}
\begin{equation}
I/(MR_s^2) = a_0 + a_1 \frac{M}{R_s} + a_4 \left(\frac{M}{R_s}\right)^4 ,
\label{IC1}
\end{equation}
\begin{equation}
I/M^3 = b_1 \left(\frac{M}{R_s}\right)^{-1} + b_2 \left(\frac{M}{R_s}\right)^{-2} 
+b_3 \left(\frac{M}{R_s}\right)^{-3} +b_4 \left(\frac{M}{R_s}\right)^{-4} .
\label{IC2}
\end{equation}
In STT these $I$-$\cal C$ relations have been considered by
Staykov et al.~\cite{Staykov:2016mbt},
employing six purely nuclear EOSs 
(SLy \cite{Douchin:2001sv}, APR4 \cite{Akmal:1998cf},
FPS \cite{Lorenz:1992zz}, GCP \cite{Goriely:2010bm},
Shen \cite{Shen:1998gq,Shen:1998by} and WFF2 \cite{Wiringa:1988tp})
and two quark EOSs
(SQSB40 \cite{GondekRosinska:2008nf} and SQSB60 \cite{GondekRosinska:2008nf}).
We here extend their study to our full set of 14 EOSs,
which, in particular, include the classes of
hyperon and hybrid EOSs, not studied before in this context.

\begin{figure}
     \begin{subfigure}[b]{0.45\textwidth}
\includegraphics[width=92mm,scale=1,angle=-90]{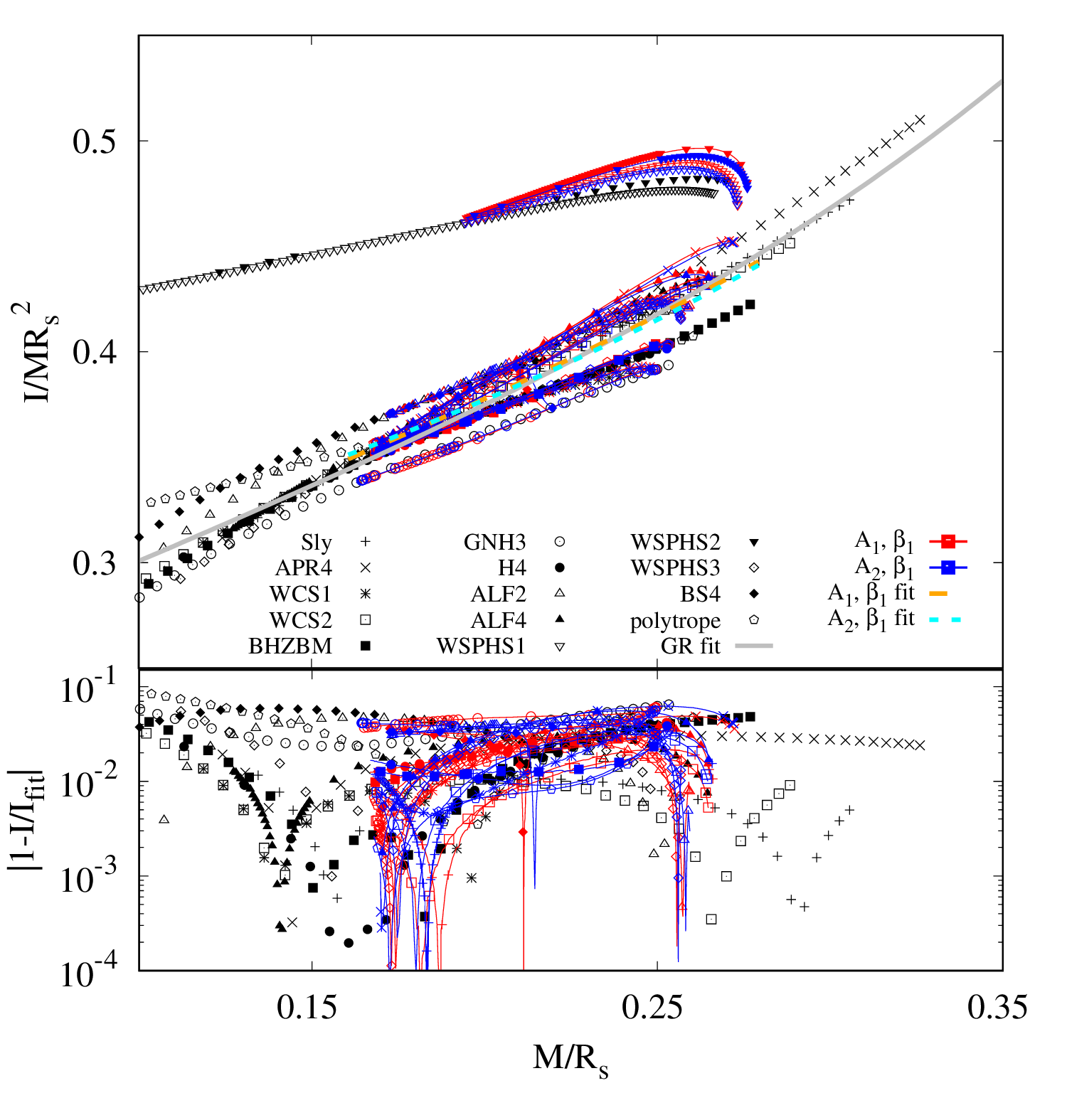}
         \caption{}
         \label{plot_sIcomp_all:a}
     \end{subfigure}
\hspace{0.7cm}
     \begin{subfigure}[b]{0.45\textwidth}
\includegraphics[width=92mm,scale=1,angle=-90]{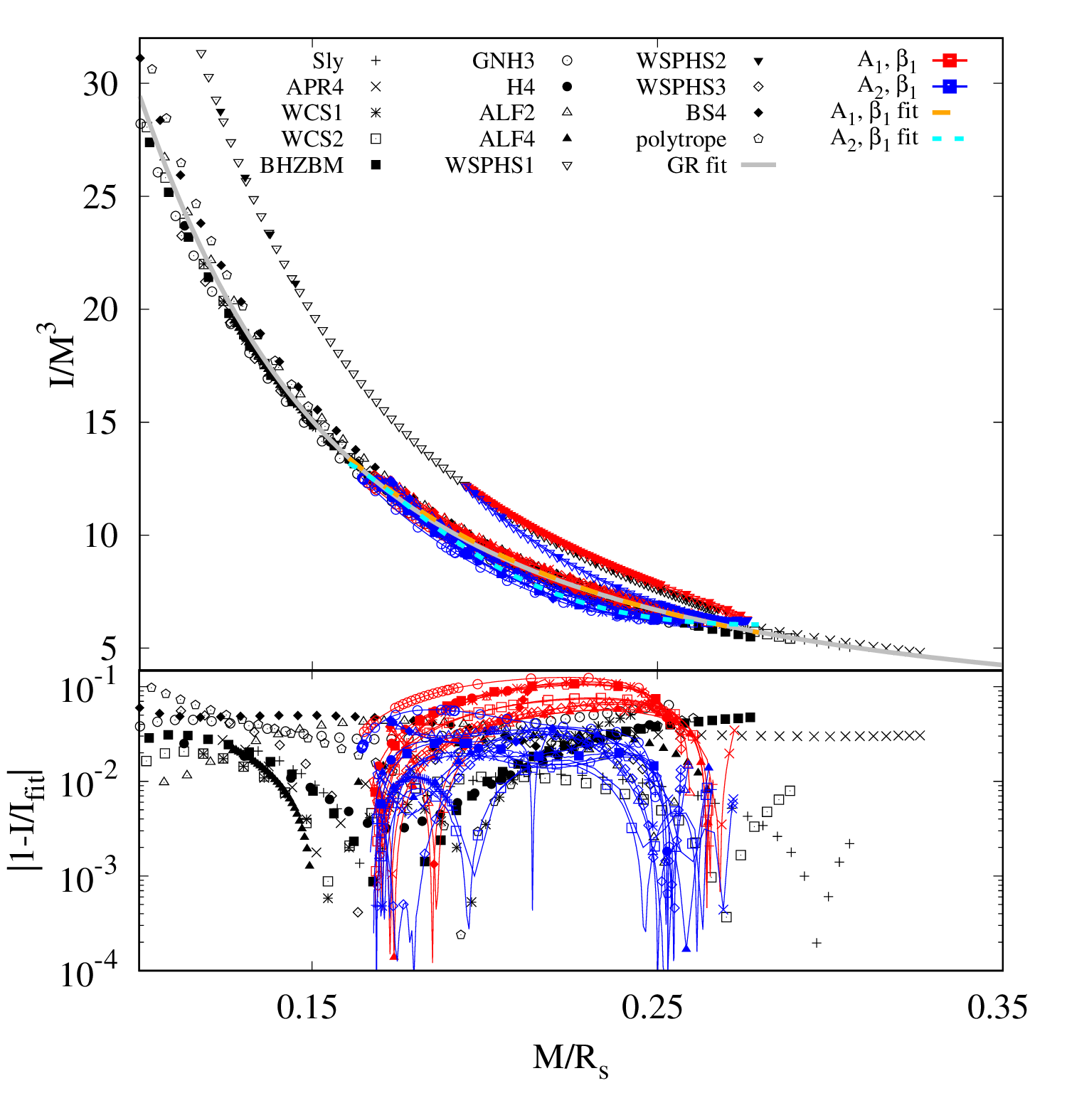}
         \caption{}
         \label{plot_sIcomp_all:b}
     \end{subfigure}
\caption{Moment of inertia $I$ versus the compactness ${\cal C}=M/R_s$
in the slow rotation approximation
for two different normalizations, $I/M R_s^2$ (a) and $I/M^3$ (b),
for all 14 EOSs, for both coupling functions $A_i$,
as well as for GR.
The upper panels show the scaled values of $I$ (symbols)
together with the fitted curves (lines) of the universal relations 
(excluding the two quark EOSs).
The lower panels exhibit the deviations
from the fitted values, $|1 - I/I_{\rm fit}|$. 
}
     \label{plot_sIcomp_all}
\end{figure}

In Fig.~\ref{plot_sIcomp_all} we present the moment of inertia $I$
as a function of the compactness $\cal C$, 
employing the two scalings $I/M R_s^2$ (a) and $I/M^3$ (b)
for all 14 EOSs. The figures include the values 
for both coupling functions $A_i$ with $\beta_1$,
as well as the GR values.
The symbols in the figures denote the respective scaled values
of $I$ versus $\cal C$, associated with the various EOSs.
The colors of the symbols mark these values in the respective theories,
i.e., GR (black), STT $A_1$, $\beta_1$ (red), STT $A_2$, $\beta_1$ (blue).

Besides the symbols associated with the various EOSs
for the scaled values of $I$,
the upper panels also show
the fitted universal relations (\ref{IC1}) and (\ref{IC2})
as solid lines:
GR (grey), STT $A_1$, $\beta_1$ (orange), STT $A_2$, $\beta_1$ (cyan).
Note, that for the GR case we have included only configurations up to the maximum mass.
We have not included the pure quark stars (WSPHS1 and WSPHS2) in the fits,
since quark stars exhibit a somewhat different behaviour \cite{Staykov:2016mbt}.
The lower panels exhibit the deviations
from the fitted values, $|1 - I/I_{\rm fit}|$,
which are always below $10\%$.

\begin{table}
\centering
\begin{tabular}{||c|c|c|c||}
    \cline{2-4}
    \multicolumn{1}{c||}{} 
     & \textbf{GR} & $\boldsymbol{A_1}$, $\boldsymbol{\beta_1}$ & $\boldsymbol{A_2}$, $\boldsymbol{\beta_1}$ \\ 
\hline\hline 
$\boldsymbol{a_0}$ & $0.232$ & $0.243$ & $0.259$ \\ 
\hline 
$\boldsymbol{a_1}$ & $0.684$ & $0.651$ & $0.553$ \\ 
\hline 
$\boldsymbol{a_4}$ & $3.813$ & $3.015$ & $4.482$ \\ 
\hline 
$\boldsymbol{\chi^2}$ & $1.13 \cdot 10^{-4}$ & $0.84 \cdot 10^{-4}$ & $0.78 \cdot 10^{-4}$ \\ 
\hline 
\hline\hline 
$\boldsymbol{b_1}$ & $1.437$ & $2.221$ & $15.115$ \\ 
\hline 
$\boldsymbol{b_2}$ & $-0.112$ & $-0.679$ & $-8.453$ \\ 
\hline 
$\boldsymbol{b_3}$ & $0.0533$ & $0.185$ & $1.707$ \\ 
\hline 
$\boldsymbol{b_4}$ & $-0.00271$ & $-0.0125$ & $-0.110$ \\ 
\hline 
$\boldsymbol{\chi^2}$ & $0.209$ & $0.0394$ & $0.0337$ \\ 
\hline 
\end{tabular} 
\caption{Fit parameters for the universal relations $I/(MR_s^2) = a_0 + a_1 u + a_4 u^4$ and  
$I/M^3 = b_1/u + b_2/u^2 +b_3/u^3+b_4/u^4$, $u=M/R_s$, 
including all EOSs except for the quark EOSs WSPHS1 and WSPHS2.}
\label{tab:IC12}
\end{table}

We exhibit the fitted coefficients for both universal relations
(\ref{IC1}) and (\ref{IC2})
in 
Table \ref{tab:IC12}
.
We find excellent agreement with the 
results of Staykov et al.~\cite{Staykov:2016mbt}.
Interestingly, the inclusion of the hyperon and hybrid EOS classes
has little impact on these universal relations for the nuclear
matter. Only pure quark matter is distinctly different.

\section{Conclusions} \label{sec:Conclusions}

We have investigated the effect of scalarization
on neutron star models with a wide variety of realistic
EOSs, including stars consisting of nucleons (3),
of nucleons and hyperons (5), of nucleons and quarks (4), 
and only of quarks (2), thus extending earlier
investigations 
\cite{Damour:1993hw,Damour:1996ke,Harada:1998ge,Salgado:1998sg,Sotani:2012eb,Pani:2014jra,Sotani:2017pfj,Doneva:2013qva,Doneva:2014uma,Doneva:2014faa,Staykov:2016mbt}
by  also considering the classes of hyperon and hybrid stars.

Restricting to static and slowly rotating models,
we have focussed on the discussion 
of the onset and the magnitude of the scalarization,
searching for its universal features.
Clearly, the compactness of the solutions is a major
component in our understanding of the phenomenon
of scalarization \cite{Damour:1993hw,Damour:1996ke},
and compactness features prominently in various
model-independent relations \cite{Yagi:2016bkt,Doneva:2017}.
In particular, we have confirmed and extended
the results of the universal $I$-$\cal C$ relations
\cite{Lattimer:2004nj,Breu:2016ufb,Staykov:2016mbt}.

However, the most striking universal feature found relates
the gravitational potential at the center of the star,
as embodied in $g_{tt}(0)$,
to the properties of the scalar field.
The scalar charge $\omega_A$, the value $\varphi_0$ of the scalar field at the
center of the star and the value $\varphi_s$ of the scalar field
at the surface of the star are all determined 
(with only a very small variance) by $g_{tt}(0)$.
This holds for all EOSs, including the quark EOSs.
Only the EOS BS4 starts to deviate from this strong
correlation close to the maximal densities, where it is known.
Indeed, the correlation is so strong, that
this exceptional deviation of the EOS BS4 suggests
that the calculations are reaching beyond the validity of this EOS,
when the deviations arise.

To support our conclusions, we have considered the effect of
scalarization not only on the widely used exponential
coupling function $A_1$ with two values of the coupling parameter
$\beta$, but also on an alternative coupling function $A_2$,
based on the hyperbolic cosine. Clearly, the onset of the
scalarization is only determined by $\beta$, while the
magnitude of scalarization is also governed by the 
coupling function, leading to less scalarization for $A_2$, 
as expected according to previous work with a 
coupling function $A_3$, based on the cosine  
\cite{Damour:1993hw,Damour:1996ke}.

Doneva et al.~\cite{Doneva:2013qva,Doneva:2014uma,Doneva:2014faa,Staykov:2016mbt}
have also studied rapidly rotating neutron stars in STT,
investigating, in particular, universal relations.
Whereas the STT results for the universal $I$-$\cal C$ relations 
do not show significant deviations from the GR results for slow rotation,
in the case of rapid rotation major deviations from GR
can occur \cite{Staykov:2016mbt}.
The group has also addressed the effect of a mass term for the scalar field
\cite{Yazadjiev:2016pcb,Doneva:2016xmf}.
In both cases the effect of scalarization is enhanced.
It will be interesting to see, whether the strong correlation
of the scalarization with the gravitational potential
is retained in the presence of rapid rotation
and for a massive scalar field.
We expect, that such a strong correlation could be present
for fixed values of the scaled angular momentum $j=J/M^2$, since $j$
has also served as an adequate ingredient in other universal relations.

Finally we would like to mention, that we have started to investigate
the presence of this correlation also for scalarized boson stars
\cite{Kleihaus:2015iea}.
Interestingly, for non-rotating boson stars (with quartic potential) 
the onset of the scalarization 
arises at almost the same value as for neutron stars,
i.e., at $\beta= -4.363$ for the boson stars of \cite{Kleihaus:2015iea} with $\Lambda=300$.
Moreover, the dependence of the scalar field charge 
on the gravitational potential ($g_{tt}(0)$) is rather similar
to the neutron star case exhibited here, leading (for fixed $\beta$)
basically to increasing concentric curves with increasing angular momentum.

\section*{Acknowledgment}

We would like to acknowledge support by the DFG Research Training Group 1620
{\sl Models of Gravity} as well as by FP7, Marie Curie Actions, People, 
International Research Staff Exchange Scheme (IRSES-606096).
BK
gratefully acknowledges support from Fundamental Research in Natural Sciences
by the Ministry of Education and Science of Kazakhstan.

\section*{Appendix: Calculation of the bifurcation points}

\subsection*{Perturbative treatment of the bifurcations}

Let us consider the case of a small scalar function $\varphi$, i.e.,
we may neglect terms of order $\varphi^2$. 
The differential equations then reduce to the equations in GR 
plus a linear equation for the scalar field,
\begin{equation}
\frac{d}{dr}\left(\sqrt{N}e^{f/2} r^2 \frac{d\varphi}{dr} \right) = 
\beta 4\pi r^2\frac{e^{f/2}}{\sqrt{N}}\left(\tilde{\varepsilon}-3 \tilde{p}\right) \varphi \ .
\label{eqphi_r}
\end{equation}
Using $ r= R\frac{x}{1-x}$ and $r^2 \frac{d}{dr} = R x^2 \frac{d}{dx}$,
we find
\begin{equation}
\frac{d}{dx}\left(\sqrt{N}e^{f/2} x^2 \frac{d\varphi}{dx} \right) = 
\beta 4\pi R^2 \frac{x^2}{(1-x)^4}
\frac{e^{f/2}}{\sqrt{N}}\left(\tilde\varepsilon-3 \tilde p\right) \varphi \ .
\label{eqphi_x}
\end{equation}
We note that the boundary conditions $\varphi\rvert_{x=1}=0$ 
and $\frac{d\varphi}{dx}\rvert_{x=0}=0$ have to be supplemented with
an additional condition to guarantee a non-trivial solution.
Since the ODE is linear we can choose without loss of generality
\mbox{$\varphi(0)=1$}.

In the following we derive a simple iteration scheme to find solutions.
For simplicity we write Eq.~(\ref{eqphi_x}) in the form
\begin{equation}
\frac{d}{dx}\left( h(x) \frac{d}{dx}\varphi\right) = - \beta V(x) \varphi(x)
\label{eqphi_p}
\end{equation}
with
\begin{equation}
 h(x) = \sqrt{N}e^{f/2} x^2 \, \ \ \ {\rm and } \ \ 
 V(x)=-4\pi R^2 \frac{x^2}{(1-x)^4}
\frac{e^{f/2}}{\sqrt{N}}\left(\tilde\varepsilon-3 \tilde p\right) \ .
\label{h_and_V}
\end{equation}
Integration then yields
\begin{equation}
h(x) \frac{d}{dx}\varphi = -\beta\int_0^x  V(x') \varphi(x') dx' \ ,
\label{step1}
\end{equation}
where the integration constant has been set to zero to ensure $\varphi'(0)=0$.
A second integration yields
\begin{equation}
\varphi(x)= -\beta\int_0^x \frac{1}{h(x')}
\left(\int_0^{x'}  V(x'') \varphi(x'') dx''\right)dx' + \beta c_0 \ .
\label{step2}
\end{equation}
The integration constant $c_0$ is determined from the boundary condition
$\varphi(1)=0$, i.e.,
\begin{equation}
c_0=\int_0^1 \frac{1}{h(x)}
\left(\int_0^{x}  V(x') \varphi(x') dx'\right)dx \ .
\label{det_c0}
\end{equation}
This leads to
\begin{equation}
\varphi(x)= -\beta\left(\int_0^x \frac{1}{h(x')}
\left(\int_0^{x'}  V(x'') \varphi(x'') dx''\right)dx' - 
\int_0^1 \frac{1}{h(x)}\left(\int_0^{x}  V(x') \varphi(x') dx'\right)dx
\right)
 \ .
\label{step3}
\end{equation}
Note that this is an implicit equation, since $\varphi$ appears on the 
rhs and the lhs of the equation.
Evaluating Eq.~(\ref{step3}) at $x=0$ yields
\begin{equation}
\varphi(0)=\beta
\int_0^1 \frac{1}{h(x)}\left(\int_0^{x}  V(x') \varphi(x') dx'\right)dx \ .
\label{det_b}
\end{equation}
Since we require $\varphi(0)=1$, we have to consider $\beta$ as a dependent
quantity. Solutions of Eq.~(\ref{step3}) exist only for certain values of
$\beta$.

The iteration scheme is now given by
\begin{eqnarray}
\varphi^{(i+1)}(x) & = & -\beta^{(i)}\left(\int_0^x \frac{1}{h(x')}
\left(\int_0^{x'}  V(x'') \varphi^{(i)}(x'') dx''\right)dx' - 
\int_0^1 \frac{1}{h(x)}\left(\int_0^{x}  V(x') \varphi^{(i)}(x') dx'\right)dx
\right) \ ,
\nonumber \\
\beta^{(i)} & = &
\left\{\int_0^1 \frac{1}{h(x)}
          \left(\int_0^{x}  V(x') \varphi^{(i)}(x') dx'\right)dx\right\}^{-1} \ .
\label{iter}
\end{eqnarray}
The iterations converge very fast. Typically 10 steps are sufficient
to determine $\beta$ up to 10 digits.

This method yields for any neutron star solution in GR the value of $\beta$ where
the bifurcation of the scalarized neutron star solutions occurs.
If the sequence of neutron star solutions in GR is characterized
by the central pressure $\tilde p_0$, one obtains $\beta_{\rm bif}(\tilde p_0)$.
The maximum of $\beta_{\rm bif}(\tilde p_0)$ in turn determines the critical
value $\beta_{\rm cr}$ beyond which no scalarized neutron star solutions exist. 
The critical $\beta_{\rm cr}$ then only depends on the EOS.

\begin{figure}[t!]
\begin{center}
\mbox{\hspace{0.2cm}
\includegraphics[height=.45\textheight, angle =-90]{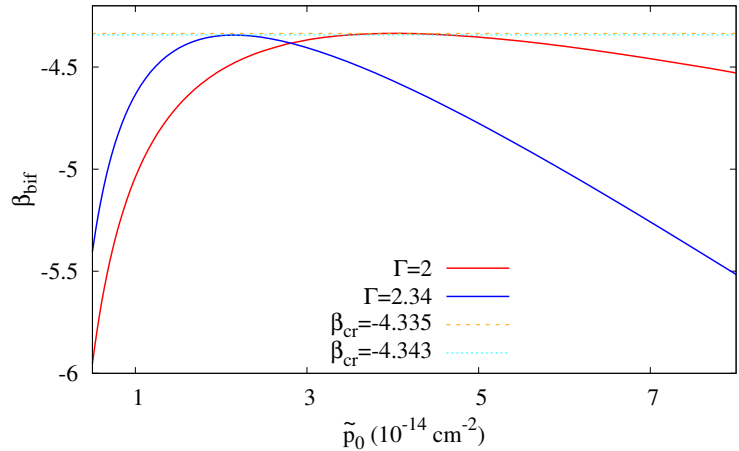}
\label{Fig_app1}
}
\end{center}
\vspace{-0.5cm}
\caption{The value of $\beta$ at the bifurcation versus
the central pressure $\tilde p_0$ for two polytropic EOSs. 
The respective critical values $\beta_{\rm cr}$ 
are indicated by horizontal lines.
\label{Fig_app}
}
\end{figure}

To demonstrate this property we consider (for simplicity)
two polytropic EOSs (\ref{poly}) 
with $K=1186.0$ and $\Gamma = 2.34$, $\beta_{\rm cr}=-4.343$  
and with $K=72.5216$ and $\Gamma = 2.0$, $\beta_{\rm cr}=-4.335$.
Fig.~\ref{Fig_app} shows $\beta_{\rm bif}$ as a function 
of the central pressure $\tilde p_0$. 
The maximal value of $\beta_{\rm bif}$ represents
the critical value $\beta_{\rm cr}$ for a given EOS.
For the two EOSs considered, these critical values
are indicated by the horizontal lines. 

\subsection*{Bifurcation points for realistic EOSs}

In order to obtain the bifurcation points for realistic EOSs using the numerical approach described in section \ref{sec:Numerical Method}, 
we implement 
the following procedure:
instead of iterating the integral relation (\ref{iter}), 
we solve the differential equation (\ref{eqphi_x}), 
but now assuming that $\beta=B(x)$ is a function of $x$
\begin{equation}
\frac{d}{dx}\left(\sqrt{N}e^{f/2} x^2 \frac{d\varphi}{dx} \right) = 
B(x) 4\pi R^2 \frac{x^2}{(1-x)^4}
\frac{e^{f/2}}{\sqrt{N}}\left(\tilde\varepsilon-3 \tilde p\right)\varphi  .
\label{eqphi_x_2}
\end{equation}
We thus add the following auxiliary differential equation to the system
\begin{equation}
\frac{d}{dx}B(x) = 0 .
\label{eqB_aux}
\end{equation}

This method is equivalent to the previous one.
We have to supply three boundary conditions, 
which we choose to be the same as described in the previous method,
$\varphi\rvert_{x=1}=0$, $\frac{d\varphi}{dx}\rvert_{x=0}=0$ and $\varphi\rvert_{x=0}=1$. 
This is implemented in COLSYS, 
together with a routine that interpolates a previously generated static solution. 
Static solutions with $500-1000$ points give good results. 
The method converges fast and works for all EOSs considered. 
The function $B(x)$ converges to a constant value, 
that determines the bifurcation point, $B(x)=\beta_{\rm bif}$. 
In principle, $\beta_{\rm bif}$ depends on the EOS 
and the central pressure of the static configuration.
This approach is similar to the method used by Harada in \cite{Harada:1997mr} to calculate the bifurcation point using quasinormal modes, when constraining to the case of vanishing imaginary part of the fequencies.  

\begin{figure}[t!]
\begin{center}
\mbox{\hspace{0.2cm}
\includegraphics[height=.45\textheight, angle =-90]{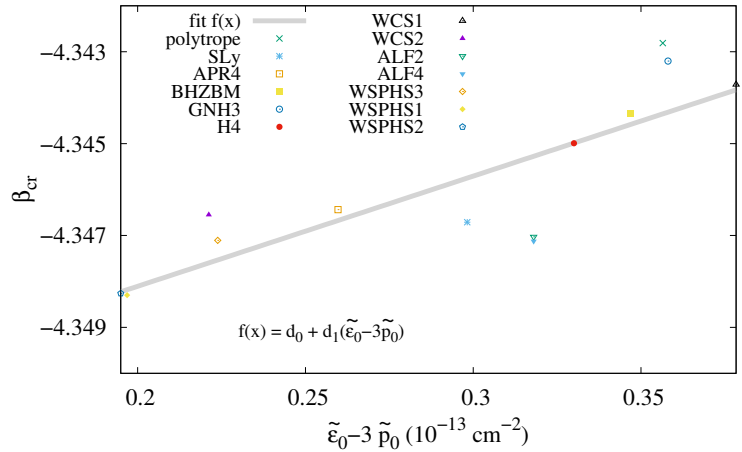}
}
\end{center}
\vspace{-0.5cm}
\caption{The critical value $\beta_{\rm cr}$ versus
the trace of the energy-momentum tensor $\tilde\varepsilon-3 \tilde p_0$, 
for all EOSs considered (except BS4). 
The grey curve is a fit to the function $f=d_0 + d_1 \left(\tilde\varepsilon-3 \tilde p_0\right)$, with $d_0=-4.353$, $d_1=0.024$ and reduced $\chi^2$ of $1.13 \cdot 10^{-6}$.
}
\label{Fig_beta_cr_rho0-3p0}
\end{figure}

The critical value $\beta_{\rm cr}$ is calculated as the maximum of $\beta_{\rm bif}$. 
In the Figures \ref{Fig_beta_cr_MR} and \ref{Fig_beta_cr_rho0-3p0}, 
the critical value $\beta_{\rm cr}$ is shown as a function of the compactness 
and of the trace of the energy-momentum tensor,
i.e., the quantity $\tilde\varepsilon-3 \tilde p_0$, respectively. 
All realistic EOSs possess very similar values of $\beta_{\rm cr}$, 
which may be related to the fact that all of them have 
similar values of the compactness ${\cal C}=M/R$ and 
the gravitational potential at the center as represented by $g_{tt}(0)$.

\end{document}